\documentclass[twocolumn,twoside,slac_two]{revtex4}
\usepackage{graphicx}
\usepackage{fancyhdr}
\begin{document}

\title{Spectroscopy and Strong Decays of Charmed Baryons}

%

\author{Hai-Yang Cheng}
\affiliation{Institute of Physics, Academia Sinica, Taipei, Taiwan
115, Republic of China}

\begin{abstract}
Spectroscopy and strong decays of the charmed baryons are reviewed.
Possible spin-parity quantum numbers of several newly observed
charmed baryon resonances are discussed. Strong decays of charmed
baryons are analyzed in the framework of heavy hadron chiral
perturbation theory in which heavy quark symmetry and chiral
symmetry are synthesized.

\end{abstract}

\maketitle

\thispagestyle{fancy}


\section{Introduction}
In the past years many new excited charmed baryon states have been
discovered by BaBar, Belle and CLEO. In particular, $B$ factories
have provided a very rich source of charmed baryons both from $B$
decays and from the continuum $e^+e^-\to c\bar c$. A new era for
the charmed baryon spectroscopy is opened by the rich mass
spectrum and the relatively narrow widths of the excited states.
Experimentally and theoretically, it is important to identify the
quantum numbers of these new states and understand their
properties. Since the pseudoscalar mesons involved in the strong
decays of charmed baryons are soft, the charmed baryon system
offers an excellent ground for testing the ideas and predictions
of heavy quark symmetry of the heavy quarks and chiral symmetry of
the light quarks.

\section{Spectroscopy}
Charmed baryon spectroscopy provides an ideal place for studying
the dynamics of the light quarks in the environment of a heavy
quark. The charmed baryon of interest contains a charmed quark and
two light quarks, which we will often refer to as a diquark. Each
light quark is a triplet of the flavor SU(3). Since ${\bf 3}\times
{\bf 3}={\bf \bar 3}+{\bf 6}$, there are two different SU(3)
multiplets of charmed baryons: a symmetric sextet {\bf 6} and an
antisymmetric antitriplet ${\bf \bar 3}$.

In the quark model, the orbital angular momentum of the light
diquark can be decomposed into ${\bf L}_\ell={\bf L}_\rho+{\bf
L}_\lambda$. where ${\bf L}_\rho$ is the orbital angular momentum
between the two light quarks and ${\bf L}_\lambda$ the orbital
angular momentum between the diquark and the charmed quark. The
lowest-lying orbitally excited baryon states are the $p$-wave
charmed baryons. Denoting the quantum numbers $L_\rho$ and
$L_\lambda$ as the eigenvalues of ${\bf L}_\rho^2$ and ${\bf
L}_\lambda^2$, respectively, the $p$-wave heavy baryon can be
either in the $(L_\rho=0,L_\lambda=1)$ $\lambda$-state or the
$(L_\rho=1,L_\lambda=0)$ $\rho$-state. It is obvious that the
orbital $\lambda$-state ($\rho$-state) is symmetric (antisymmetric)
under the interchange of two light quarks $q_1$ and $q_2$. The
total angular momentum of the diquark is ${\bf J}_\ell={\bf
S}_\ell+{\bf L}_\ell$ and the total angular momentum of the charmed
baryon is ${\bf J}={\bf S}_c+{\bf J}_\ell$. In the heavy quark
limit, the spin of the charmed quark $S_c$ and the total angular
momentum of the two light quarks $J_\ell$ are separately conserved.

There are seven lowest-lying $p$-wave $\Lambda_c$ arising from
combining the charmed quark spin $S_c$ with light constituents in
$J_\ell^{P_\ell}=1^-$ state: three $J^P=\frac12^-$ states, three
$J^P=\frac32^-$ states and one $J^P=\frac52^-$ state. They form
three doublets $\Lambda_{c1}({1\over 2}^-,{3\over
2}^-),\tilde\Lambda_{c1}({1\over 2}^-,{3\over
2}^-),\tilde\Lambda_{c2}({3\over 2}^-,{5\over 2}^-)$ and one
singlet $\tilde\Lambda_{c0}({1\over 2}^-)$ in the notation
$\Lambda_{cJ_\ell}(J^P)$, where we have used a tilde to denote the
multiplets antisymmetric in the orbital wave functions under the
exchange of two light quarks. Quark models \cite{Capstick} indicate
that the untilde states for $\Lambda$- and $\Sigma$-type charmed
baryons with symmetric orbital wave functions lie about 150 MeV
below the tilde ones. The two states in each doublet with
$J=J_\ell\pm{1\over 2}$ are nearly degenerate; their masses split
only by a chromomagnetic interaction.

The next orbitally excited states are the positive parity
excitations with $L_\rho+L_\lambda=2$.  There are two multiplets
for the first positive-parity excited $\Lambda_c$ with the
symmetric orbital wave function, corresponding to
$L_\lambda=2,L_\rho=0,L=2$ and $L_\lambda=0,L_\rho=2,L=2$, see
Table \ref{tab:pp} (for other charmed baryons, see \cite{CC} for
details).
For the case of $L_\lambda=L_\rho=1$, the total orbital angular
momentum $L_\ell$ of the diquark is 2, 1 or 0. Since the orbital
states are antisymmetric under the interchange of two light quarks,
we shall use a tilde to denote the $L_\lambda=L_\rho=1$ states.
The Fermi-Dirac statistics for baryons yields seven more multiplets
for positive-parity excited $\Lambda_c$ states.

\begin{table}[h]
\caption{The first positive-parity excitations of $\Lambda$ charmed
baryons and their quantum numbers. States with antisymmetric
orbital wave functions (i.e. $L_\rho=L_\lambda=1$) under the
interchange of two light quarks are denoted by a tilde. There are
two multiplets $\Lambda_{c2}$ and $\hat\Lambda_{c2}$ with symmetric
orbital wave functions arising from the orbital states
$L_\rho=0,L_\lambda=2$ and $L_\rho=2,L_\lambda=0$, respectively. We
use a hat to distinguish between them.} \label{tab:pp}
\begin{center}
\begin{tabular}{|c|cccc|} \hline\hline
~~~~~State~~~~~ & SU(3)$_F$ & ~~$S_\ell$~~ & ~~$L_\ell$~~&
~~$J_\ell^{P_\ell}$~~  \\
 \hline
 $\Lambda_{c2}({3\over 2}^+,{5\over 2}^+)$ & ${\bf \bar 3}$ & 0 & 2 &
 $2^+$  \\
 $\hat{\Lambda}_{c2}({3\over 2}^+,{5\over 2}^+)$ & ${\bf \bar 3}$ & 0 & 2
 & $2^+$  \\
 $\tilde\Lambda_{c1}({1\over 2}^+,\frac32^+)$ & ${\bf \bar 3}$ & 1 & 0 & $1^+$  \\
 $\tilde\Lambda'_{c0}({1\over 2}^+)$ & ${\bf \bar 3}$ & 1 & 1 & $0^+$  \\
 $\tilde\Lambda'_{c1}({1\over 2}^+,{3\over 2}^+)$ & ${\bf \bar 3}$ & 1 & 1 & $1^+$  \\
 $\tilde\Lambda'_{c2}({3\over 2}^+,{5\over 2}^+)$ & ${\bf \bar 3}$ & 1 & 1 & $2^+$  \\
 $\tilde\Lambda''_{c1}({1\over 2}^+,{3\over 2}^+)$ & ${\bf \bar 3}$ & 1 & 2 & $1^+$  \\
 $\tilde\Lambda''_{c2}({3\over 2}^+,{5\over 2}^+)$ & ${\bf \bar 3}$ & 1 & 2 & $2^+$ \\
 $\tilde\Lambda''_{c3}({5\over 2}^+,{7\over 2}^+)$ & ${\bf \bar 3}$ & 1 & 2 & $3^+$ \\
 \hline \hline
\end{tabular}
\end{center}
\end{table}

\begin{table*}[t]
\caption{Mass spectra and decay widths (in units of MeV) of charmed
baryons taken from \cite{CC}. Except for the parity of the lightest
$\Lambda_c^+$ and the spin-parity of $\Lambda_c(2880)^+$, none of
the other $J^P$ quantum numbers given in the table has been
measured. One has to rely on the quark model to determine the
spin-parity assignments.} \label{tab:spectrum}
\begin{center}
\begin{tabular}{|c|c|ccc|c|c|c|c|} \hline \hline
~~State~~ & ~~$J^P$~~ &~$S_\ell$~ & ~$L_\ell$~ &
~$J_\ell^{P_\ell}$~ &
~~~~~~~~~Mass~~~~~~~~~ & ~~~~Width~~~~ &~Principal decay modes~\\
\hline
 $\Lambda_c^+$ & ${1\over 2}^+$ & 0 & 0 & $0^+$ & $2286.46\pm0.14$ & & weak  \\
 \hline
 $\Lambda_c(2595)^+$ & ${1\over 2}^-$ & 0 & 1 & $1^-$ & $2595.4\pm0.6$ &
 $3.6^{+2.0}_{-1.3}$ & $\Sigma_c\pi,\Lambda_c\pi\pi$ \\
 \hline
 $\Lambda_c(2625)^+$ & ${3\over 2}^-$ & 0 & 1 & $1^-$ & $2628.1\pm0.6$ &
 $<1.9$ & $\Lambda_c\pi\pi,\Sigma_c\pi$ \\
 \hline
 $\Lambda_c(2765)^+$ & $?^?$ & ? & ? & $?$ & $2766.6\pm2.4$ & $50$ & $\Sigma_c\pi,\Lambda_c\pi\pi$ \\
 \hline
 $\Lambda_c(2880)^+$ & ${5\over 2}^+$ & ? & ? & ? & $2881.5\pm0.3$ & $5.5\pm0.6$
 & $\Sigma_c^{(*)}\pi,\Lambda_c\pi\pi,D^0p$ \\
 \hline
 $\Lambda_c(2940)^+$ & $?^?$ & ? & ? & $?$ & $2938.8\pm1.1$ & $13.0\pm5.0$ &
 $\Sigma_c^{(*)}\pi,\Lambda_c\pi\pi,D^0p$ \\ \hline
 $\Sigma_c(2455)^{++}$ & ${1\over 2}^+$ & 1 & 0 & $1^+$ & $2454.02\pm0.18$ &
 $2.23\pm0.30$ & $\Lambda_c\pi$ \\
 \hline
 $\Sigma_c(2455)^{+}$ & ${1\over 2}^+$ & 1 & 0 & $1^+$ & $2452.9\pm0.4$ &
 $<4.6$ & $\Lambda_c\pi$\\
 \hline
 $\Sigma_c(2455)^{0}$ & ${1\over 2}^+$ & 1 & 0 & $1^+$ & $2453.76\pm0.18$
 & $2.2\pm0.4$ & $\Lambda_c\pi$ \\
 \hline
 $\Sigma_c(2520)^{++}$ & ${3\over 2}^+$ & 1 & 0 & $1^+$ & $2518.4\pm0.6$
 & $14.9\pm1.9$ & $\Lambda_c\pi$\\
 \hline
 $\Sigma_c(2520)^{+}$ & ${3\over 2}^+$ & 1 & 0 & $1^+$ & $2517.5\pm2.3$
 & $<17$ & $\Lambda_c\pi$ \\
 \hline
 $\Sigma_c(2520)^{0}$ & ${3\over 2}^+$ & 1 & 0 & $1^+$ & $2518.0\pm0.5$
 & $16.1\pm2.1$ & $\Lambda_c\pi$ \\
 \hline
 $\Sigma_c(2800)^{++}$ & ${3\over 2}^-$? & 1 & 1 & $2^-$ & $2801^{+4}_{-6}$ & $75^{+22}_{-17}$ &
 $\Lambda_c\pi,\Sigma_c^{(*)}\pi,\Lambda_c\pi\pi$ \\
 \hline
 $\Sigma_c(2800)^{+}$ & ${3\over 2}^-$? & 1 & 1 & $2^-$ & $2792^{+14}_{-5}$ & $62^{+60}_{-40}$ &
 $\Lambda_c\pi,\Sigma_c^{(*)}\pi,\Lambda_c\pi\pi$ \\
 \hline
 $\Sigma_c(2800)^{0}$ & ${3\over 2}^-$? &1 & 1 & $2^-$ & $2802^{+4}_{-7}$ & $61^{+28}_{-18}$ &
 $\Lambda_c\pi,\Sigma_c^{(*)}\pi,\Lambda_c\pi\pi$\\
 \hline
 $\Xi_c^+$ & ${1\over 2}^+$ & 0 & 0 & $0^+$ & $2467.9\pm0.4$ & & weak \\ \hline
 $\Xi_c^0$ & ${1\over 2}^+$ & 0 & 0 & $0^+$ & $2471.0\pm0.4$ & & weak \\ \hline
 $\Xi'^+_c$ & ${1\over 2}^+$ & 1 & 0 & $1^+$ & $2575.7\pm3.1$ & & $\Xi_c\gamma$ \\ \hline
 $\Xi'^0_c$ & ${1\over 2}^+$ & 1 & 0 & $1^+$ & $2578.0\pm2.9$ & & $\Xi_c\gamma$ \\ \hline
 $\Xi_c(2645)^+$ & ${3\over 2}^+$ & 1 & 0 & $1^+$ & $2646.6\pm1.4$ & $<3.1$ & $\Xi_c\pi$ \\
 \hline
 $\Xi_c(2645)^0$ & ${3\over 2}^+$ & 1 & 0 & $1^+$ & $2646.1\pm1.2$ & $<5.5$ & $\Xi_c\pi$ \\
 \hline
 $\Xi_c(2790)^+$ & ${1\over 2}^-$ & 0 & 1 & $1^-$ & $2789.2\pm3.2$ & $<15$ & $\Xi'_c\pi$\\
 \hline
 $\Xi_c(2790)^0$ & ${1\over 2}^-$ & 0 & 1 & $1^-$ & $2791.9\pm3.3$ & $<12$ & $\Xi'_c\pi$ \\
 \hline
 $\Xi_c(2815)^+$ & ${3\over 2}^-$ & 0 & 1 & $1^-$ & $2816.5\pm1.2$ & $<3.5$ & $\Xi^*_c\pi,\Xi_c\pi\pi,\Xi_c'\pi$ \\
 \hline
 $\Xi_c(2815)^0$ & ${3\over 2}^-$ & 0 & 1 & $1^-$ & $2818.2\pm2.1$ & $<6.5$ & $\Xi^*_c\pi,\Xi_c\pi\pi,\Xi_c'\pi$ \\
 \hline
 $\Xi_c(2980)^+$ & $?^?$ & ? & ? & $?$ & $2971.1\pm1.7$ & $25.2\pm3.0$
 &  see Table 7 of \cite{CC}  \\
 \hline
 $\Xi_c(2980)^0$ & $?^?$ & ? & ? & $?$ & $2977.1\pm9.5$ & $43.5$
 &  see Table 7 of \cite{CC}  \\
 \hline
 $\Xi_c(3055)^+$ & $?^?$ & ? & ? & $?$ & $3054.2\pm1.3$ & $17\pm13$ &
   $\Lambda_c^+\bar K^0$, $\Lambda_c^+K^-\pi^+$ \\
 \hline
 $\Xi_c(3080)^+$ & $?^?$ & ? & ? & $?$ & $3076.5\pm0.6$ & $6.2\pm1.1$ &
 see Table 7 of \cite{CC}  \\
 \hline
 $\Xi_c(3080)^0$ & $?^?$ & ? & ? & $?$ & $3082.8\pm2.3$ & $5.2\pm3.6$
 &  see Table 7 of \cite{CC}  \\
 \hline
 $\Xi_c(3123)^+$ & $?^?$ & ? & ? & $?$ & $3122.9\pm1.3$ & $4.4\pm3.8$ &
   $\Lambda_c^+\bar K^0$, $\Lambda_c^+K^-\pi^+$ \\
 \hline
 $\Omega_c^0$ & ${1\over 2}^+$ & 1 & 0 & $1^+$ & $2697.5\pm2.6$ & & weak \\
 \hline
 $\Omega_c(2770)^0$ & ${3\over 2}^+$ & 1 & 0 & $1^+$ & $2768.3\pm3.0$ & & $\Omega_c\gamma$ \\
 \hline \hline
\end{tabular}
\end{center}
\end{table*}

The observed mass spectra and decay widths of charmed baryons are
summarized in Table \ref{tab:spectrum}. For the experimental status
of charmed baryons, see \cite{Mizuk}. In the following we discuss
some of the new excited charmed baryon states:
\subsection{$\Lambda_c$}

It is known that $\Lambda_c(2595)^+$ and $\Lambda_c(2625)^+$ form a
doublet $\Lambda_{c1}({1\over 2}^-,{3\over 2}^-)$ \cite{Cho}. The
dominant decay mode is $\Sigma_c\pi$ in an $S$ wave for
$\Lambda_{c1}({1\over 2}^-)$ and $\Lambda_c\pi\pi$ in a $P$ wave
for $\Lambda_{c1}({3\over 2}^-)$. (The two-body mode $\Sigma_c\pi$
is a $D$-wave in $\Lambda_c(\frac32^-)$ decay.) This explains why
the width of $\Lambda_c(2625)^+$ is narrower than that of
$\Lambda_c(2595)^+$.

$\Lambda_c(2765)^+$ is a broad state ($\Gamma\approx 50$ MeV) first
seen in $\Lambda_c^+\pi^+\pi^-$ by CLEO \cite{CLEO:Lamc2880}. It
appears to resonate through $\Sigma_c$ and probably also
$\Sigma_c^*$. However, whether it is a $\Lambda_c^+$ or a
$\Sigma_c^+$ or whether the width might be due to overlapping
states are not known. According to PDG \cite{PDG}, this state has a
nickname, namely, $\Sigma_c(2765)^+$. The Skyrme model \cite{Oh}
and the quark model \cite{Capstick} suggest a $J^P=\frac12^+$
$\Lambda_c$ state with a mass 2742 and 2775 MeV, respectively.
Therefore, $\Lambda_c(2765)^+$ could be a first positive-parity
excitation of $\Lambda_c$. However, two recent studies based on the
relativistic quark model advocate a different assignment: a radial
excitation $2{1\over 2}^+$ by \cite{Ebert} and a negative-parity
state with $J^P={5\over 2}^-$ by \cite{Ger}.

The state $\Lambda_c(2880)^+$ first observed by CLEO
\cite{CLEO:Lamc2880} in $\Lambda_c^+\pi^+\pi^-$ was also seen by
BaBar in the $D^0p$ spectrum \cite{BaBar:Lamc2940}. It was
originally conjectured that, based on its narrow width,
$\Lambda_c(2880)^+$ might be a $\tilde\Lambda^+_{c0}({1\over 2}^-)$
state \cite{CLEO:Lamc2880}. Recently, Belle has studied the
experimental constraint on the $J^P$ quantum numbers of
$\Lambda_c(2880)^+$ \cite{Belle:Lamc2880}. The angular analysis of
$\Lambda_c(2880)^+\to\Sigma_c^{0,++}\pi^\pm$ indicates that
$J=\frac52$ is favored over $J=\frac12$ or $\frac32$. In the quark
model, the candidates for the spin-${5\over 2}$ state are
$\Lambda_{c2}(\frac52^+)$, $\hat\Lambda_{c2}(\frac52^+)$,
$\tilde\Lambda_{c2}(\frac52^-)$, $\tilde\Lambda'_{c2}(\frac52^+)$,
$\tilde\Lambda''_{c2}(\frac52^+)$ and
$\tilde\Lambda''_{c3}(\frac52^+)$ (see Table \ref{tab:pp}). And
only one of them has odd parity.

Belle has also studied the resonant structure of
$\Lambda_c(2880)^+\to\Lambda_c^+\pi^+\pi^-$ and found the existence
of the $\Sigma_c^*\pi$ intermediate states \cite{Belle:Lamc2880}.
The ratio of $\Sigma_c^*\pi/\Sigma\pi$ is measured to be
 \begin{eqnarray} \label{eq:R}
 R\equiv {\Gamma(\Lambda_c(2880)\to\Sigma_c^*\pi^\pm)\over
 \Gamma(\Lambda_c(2880)\to\Sigma_c\pi^\pm)}=(24.1\pm6.4^{+1.1}_{-4.5})\%.
 \end{eqnarray}
For $J^P=\frac52^-$, $\Lambda_c(2880)$ decays to $\Sigma_c^*\pi$
and $\Sigma_c\pi$ in a $D$ wave and we obtain
 \begin{eqnarray}
 {\Gamma\left(\tilde\Lambda_{c2}(5/2^-)\to
[\Sigma_c^*\pi]_D\right)\over
\Gamma\left(\tilde\Lambda_{c2}(5/2^-)\to
[\Sigma_c\pi]_D\right)}&=&{7\over 2}
\,{p_\pi^5(\Lambda_c(2880)\to\Sigma_c^*\pi)\over
p_\pi^5(\Lambda_c(2880)\to\Sigma_c\pi)} \nonumber \\
&=&1.45\,,
 \end{eqnarray}
where the factor of 7/2 follows from heavy quark symmetry. Hence,
the assignment of $J^P={5\over 2}^-$ for $\Lambda_c(2880)$ is
disfavored. For $J^P=\frac52^+$, $\Lambda_{c2}$,
$\hat\Lambda_{c2}$, $\tilde\Lambda'_{c2}$ and
$\tilde\Lambda''_{c2}$  with $J_\ell=2$ decay to $\Sigma_c\pi$ in a
$F$ wave and $\Sigma_c^*\pi$ in $F$ and $P$ waves. Neglecting the
$P$-wave contribution for the moment,
 \begin{eqnarray}
 {\Gamma\left(\Lambda_{c2}(5/2^+)\to
[\Sigma_c^*\pi]_F\right)\over \Gamma\left(\Lambda_{c2}(5/2^+)\to
[\Sigma_c\pi]_F\right)}&=& {4\over
5}\,{p_\pi^7(\Lambda_c(2880)\to\Sigma_c^*\pi)\over
p_\pi^7(\Lambda_c(2880)\to\Sigma_c\pi)} \nonumber \\
&=& 0.23\,.
 \end{eqnarray}
At first glance, it appears that this is in good agreement with
experiment. However, the $\Sigma_c^*\pi$ channel is available via a
$P$-wave and is enhanced by a factor of $1/p_\pi^4$ relative to the
$F$-wave one. Unfortunately, we cannot apply heavy quark symmetry
to calculate the contribution of the $[\Sigma_c^*\pi]_F$ channel to
the ratio $R$ as the reduced matrix elements are different for
$P$-wave and $F$-wave modes. In this case, one has to reply on a
phenomenological model to compute the ratio $R$. At any event, the
$\Sigma_c^*\pi$ mode produced in $\Lambda_c(2880)$ is {\it a
priori} not necessarily suppressed relative to $[\Sigma_c\pi]_F$.
Therefore, if $\Lambda_c(2880)^+$ is one of the states
$\Lambda_{c2}$, $\hat\Lambda_{c2}$, $\tilde\Lambda'_{c2}$ and
$\tilde\Lambda''_{c2}$, the prediction $R=0.23$ is not robust as it
can be easily upset by the contribution from the $P$-wave
$\Sigma_c^*\pi$.

As for $\tilde\Lambda''_{c3}(\frac52^+)$, it decays to
$\Sigma_c^*\pi$, $\Sigma_c\pi$ and $\Lambda_c\pi$ all in $F$ waves.
Since $J_\ell=3,L_\ell=2$, it turns out that
 \begin{eqnarray}
 {\Gamma\left(\Lambda''_{c3}(5/2^+)\to
[\Sigma_c^*\pi]_F\right)\over \Gamma\left(\Lambda''_{c3}(5/2^+)\to
[\Sigma_c\pi]_F\right)}&=&{5\over
4}\,{p_\pi^7(\Lambda_c(2880)\to\Sigma_c^*\pi)\over
p_\pi^7(\Lambda_c(2880)\to\Sigma_c\pi)} \nonumber \\
&=& 0.36\,.
 \end{eqnarray}
Although this deviates from the experimental measurement
(\ref{eq:R}) by $1\sigma$, it is a robust prediction. This has
motivated Chun-Khiang Chua and me to conjecture that that the first
positive-parity excited charmed baryon $\Lambda_c(2880)^+$ could be
an admixture of $\Lambda_{c2}(\frac52^+)$,
$\hat\Lambda_{c2}(\frac52^+)$ and $\Lambda''_{c3}(\frac52^+)$
\cite{CC}.

It is worth mentioning that very recently the Peking group
\cite{Zhu} has studied the strong decays of charmed baryons based
on the so-called $^3P_0$ recombination model. For the
$\Lambda_c(2880)$, Peking group found that (i) the possibility of
$\Lambda_c(2880)$ being a radial excitation is ruled out as its
decay into $D^0p$ is prohibited in the $^3P_0$ model if
$\Lambda_c(2880)$ is a first radial excitation of $\Lambda_c$, and
(ii) the only possible assignment is $\Lambda''_{c3}(\frac52^+)$
since according to the $^3P_0$ model \cite{Zhu}
 \begin{eqnarray}
 {\Gamma\left(\Lambda_{c2}(5/2^+)\to
\Sigma_c^*\pi\right)\over \Gamma\left(\Lambda_{c2}(5/2^+)\to
\Sigma_c\pi\right)} &=& 0.06\,, \nonumber \\
 {\Gamma\left(\hat\Lambda_{c2}(5/2^+)\to
\Sigma_c^*\pi\right)\over \Gamma\left(\hat \Lambda_{c2}(5/2^+)\to
\Sigma_c\pi\right)} &=& 78.3\,.
 \end{eqnarray}
Both symmetric states $\Lambda_{c2}$ and $\hat\Lambda_{c2}$ are
thus ruled out as the predicted ratio $R$ is either too small or
too big compared to experiment. However, the assignment of
$\Lambda''_{c3}(\frac52^+)$ for $\Lambda_c(2880)$ has an issue with
the spectrum: The quark model indicates a $\Lambda_{c2}(\frac52^+)$
state around 2910 MeV which is close to the mass of
$\Lambda_c(2880)$, while the mass of $\Lambda''_{c3}(\frac52^+)$ is
higher \cite{Capstick}.

It is interesting to notice that, based on the diquark idea, the
quantum numbers $J^P=\frac52^+$ have been correctly predicted in
\cite{Selem} for the $\Lambda_c(2880)$ before the Belle experiment.

The highest $\Lambda_c(2940)^+$ was first discovered by BaBar in
the $D^0p$ decay mode \cite{BaBar:Lamc2940} and  confirmed by Belle
in the decays $\Sigma_c^0\pi^+,\Sigma_c^{++}\pi^-$ which
subsequently decay into $\Lambda_c^+\pi^+\pi^-$
\cite{Belle:Lamc2880}. Since the mass of $\Lambda_c(2940)^+$ is
barely below the threshold of $D^{*0}p$, this observation has
motivated the authors of \cite{He} to suggest an exotic molecular
state of $D^{*0}$ and $p$ with a binding energy of order 6 MeV and
$J^P={1\over 2}^-$ for $\Lambda_c(2940)^+$. The quark potential
model predicts a $\frac52^-$ $\Lambda_c$ state at 2900 MeV and a
$\frac32^+$ $\Lambda_c$ state at 2910 MeV \cite{Capstick}. A
similar result of 2906 MeV for $\frac32^+$ $\Lambda_c$ is also
obtained in the relativistic quark model \cite{Garcilazo}. Given
the uncertainty of order 50 MeV for the quark model calculation,
this suggests that the possible allowed $J^P$ numbers of the
highest $\Lambda_c(2940)^+$ are $\frac52^-$ and $\frac32^+$. Hence,
the potential candidates are $\tilde\Lambda_{c2}(\frac52^-)$,
$\Lambda_{c2}(\frac32^+)$, $\hat\Lambda_{c2}(\frac32^+)$,
$\tilde\Lambda'_{c1}(\frac32^+)$, $\tilde\Lambda''_{c1}(\frac32^+)$
and $\tilde\Lambda''_{c2}(\frac32^+)$.  Since the predicted ratios
differ significantly for different $J^P$ quantum numbers, the
measurements of the ratio of $\Sigma_c^*\pi/\Sigma_c\pi$ will
enable us to discriminate the $J^P$ assignments for
$\Lambda_c(2940)$ \cite{CC}. Note that it has been argued in
\cite{Ebert} that $\Lambda_c(2940)$ is the first radial excitation
of $\Sigma_c$ (not $\Lambda_c$ !) with $J^P=3/2^+$.

\subsection{$\Sigma_c$}
The highest isotriplet charmed baryons $\Sigma_c(2800)^{++,+,0}$
decaying to $\Lambda_c^+\pi$ were first measured by Belle
\cite{Belle:Sigc2800}.  They are most likely to be the
$J^P=\frac32^-$ $\Sigma_{c2}$ states because the
$\Sigma_{c2}({3\over 2}^-)$ baryon decays principally into the
$\Lambda_c\pi$ system in a $D$-wave, while $\Sigma_{c1}({3\over
2}^-)$ decays mainly to the two pion system $\Lambda_c\pi\pi$ in a
$P$-wave. The state $\Sigma_{c0}({1\over 2}^-)$ can decay into
$\Lambda_c\pi$ in an $S$-wave, but it is very broad with width of
order 406 MeV. Therefore, $\Sigma_c(2800)^{++,+,0}$ are likely to
be $\Sigma_{c2}({3\over 2}^-)$ with a possible small mixing with
$\Sigma_{c0}({1\over 2}^-)$.

\subsection{$\Xi_c$}

The states $\Xi_c(2790)$ and $\Xi_c(2815)$ form a doublet
$\Xi_{c1}({1\over 2}^-,{3\over 2}^-)$. Since the diquark transition
$1^-\to 0^++\pi$ is prohibited, $\Xi_{c1}({1\over 2}^-,{3\over
2}^-)$ cannot decay to $\Xi_c\pi$. The dominant decay mode is
$[\Xi'_c\pi]_S$ for $\Xi_{c1}({1\over 2}^-)$ and $[\Xi_c^*\pi]_S$
for $\Xi_{c1}({3\over 2}^-)$ where $\Xi_c^*$ stands for
$\Xi_c(2645)$.

The new charmed strange baryons $\Xi_c(2980)^+$ and $\Xi_c(3080)^+$
that decay into $\Lambda_c^+K^-\pi^+$ were first observed by Belle
\cite{Belle:Xic2980} and confirmed by BaBar \cite{BaBar:Xic2980}.
For the charmed states $\Xi_c(2980)$ and $\Xi_c(3080)$, they could
be the first positive-parity excitations of $\Xi_c$ in viewing of
their large masses. Since the mass difference between the
antitriplets $\Lambda_c$ and $\Xi_c$ for
$J^P=\frac12^+,\frac12^-,\frac32^-$ is of order $180\sim 200$ MeV,
it is conceivable that $\Xi_c(2980)$ and $\Xi_c(3080)$ are the
counterparts of $\Lambda_c(2765)$ and $\Lambda_c(2880)$,
respectively, in the strange charmed baryon sector. As noted in
passing, the state $\Lambda_c(2765)^+$ could be an even-parity
orbital excitation or a radial excitation and $\Lambda_c(2880)$ has
the quantum numbers $J^P={5\over 2}^+$, it is thus tempting to
assign $J^P=1\frac12^+$  for $\Xi_c(2980)$ and $\frac52^+$ for
$\Xi_c(3080)$. The possible strong decays of the first
positive-parity excitations of the $\Xi_c$ states are summarized in
Table VII of \cite{CC}. Since the two-body modes $\Xi_c\pi$,
$\Lambda_cK$, $\Xi'_c\pi$ and $\Sigma_cK$ are in $P$ ($F$) waves
and the three-body modes $\Xi_c\pi\pi$ and $\Lambda_cK\pi$ are in
$S$ ($D$) waves in the decays of $\frac12^+$ ($\frac52^+$), this
explains why $\Xi_c(2980)$ is broader than $\Xi_c(3080)$. Since
both $\Xi_c(2980)$ and  $\Xi_c(3080)$ are above the $D\Lambda$
threshold, it is important to search for them in the $D\Lambda$
spectrum as well.

Two new $\Xi_c$ resonances $\Xi_c(3055)$ and $\Xi_c(3123)$ were
recently reported by BaBar \cite{BaBar:Xic3055} with masses and
widths shown in Table \ref{tab:spectrum}.

\subsection{$\Omega_c$}
At last, the $J^P=\frac32^+$ $\Omega_c(2770)$ charmed baryon was
recently observed by BaBar in the decay
$\Omega_c(2770)^0\to\Omega_c^0\gamma$ \cite{BaBar:Omegacst}. With
this new observation, the $\frac32^+$ sextet is finally completed.
However, it will be very difficult to measure the electromagnetic
decay rate  because the width of $\Omega_c^*$, which is predicted
to be of order 0.9 keV \cite{ChengSU3}, is too narrow to be
experimentally resolvable.\\

The possible spin-parity quantum numbers of the newly discovered
charmed baryon resonances that have been suggested in the
literature are summarized in Table \ref{tab:qn}. Some of the
predictions are already ruled out by experiment. For example,
$\Lambda_c(2880)$ has $J^P={5\over 2}^+$ as seen by Belle.
Certainly, more experimental studies are needed in order to pin
down the quantum numbers.

\begin{table*}[t]
\caption{Possible spin-parity quantum numbers for the newly
discovered charmed baryon resonances that have been proposed in the
literature. First radial excited states are denoted by $2J^P$.}
\label{tab:qn}
\begin{center}
\begin{tabular}{|l|c|c| c| c| c| c|} \hline \hline
  &  $\Lambda_c(2765)$ & $\Lambda_c(2880)$ & $\Lambda_c(2940)$ & $\Sigma_c(2800)$
 & $\Xi_c(2980)$ &  $\Xi_c(3080)$ \\  \hline
 Ebert et al. \cite{Ebert} & $2{1\over 2}^+,{3\over 2}^-(\Sigma_c)$ & ${5\over 2}^+$ & $2{3\over
 2}^+(\Sigma_c)$ &  ${1\over 2}^-,{3\over 2}^-,{5\over 2}^-$ & $2{1\over 2}^+$ & ${5\over 2}^+$ \\
 Garcilazo et al. \cite{Garcilazo} & ${1\over 2}^+$ & ~${1\over
 2}^-,~{3\over 2}^-$ & ${3\over 2}^+$ & ${1\over 2}^-,{3\over
 2}^-$& & \\
 Gerasyuata et al. \cite{Ger} & ${5\over 2}^-$ & ${1\over 2}^-$ & & ${5\over 2}^-$ & &  \\
 Capstick et al. \cite{Capstick} & ${1\over 2}^+$ & & ${3\over
 2}^+,{5\over 2}^-$ & & & \\
 Cheng et al. \cite{CC} & & & & & ${1\over 2}^+$ & ${5\over 2}^+$ \\
 Wilczek et al. \cite{Selem} & & ${5\over 2}^+$ & & & & \\
 He et al. \cite{He} & & & ${1\over 2}^-$ & & & \\
 \hline \hline
\end{tabular}
\end{center}
\end{table*}

\section{Strong decays}
Due to the rich mass spectrum and the relatively narrow widths of
the excited states, the charmed baryon system offers an excellent
ground for testing the ideas and predictions of heavy quark
symmetry and light flavor SU(3) symmetry. The pseudoscalar mesons
involved in the strong decays of charmed baryons such as
$\Sigma_c\to\Lambda_c\pi$ are soft. Therefore, heavy quark symmetry
of the heavy quark and chiral symmetry of the light quarks will
have interesting implications for the low-energy dynamics of heavy
baryons interacting with the Goldstone bosons.

The strong decays of charmed baryons are most conveniently
described by the heavy hadron chiral Lagrangians in which heavy
quark symmetry and chiral symmetry are incorporated
\cite{Yan,Wise}. The Lagrangian involves two coupling constants
$g_1$ and $g_2$ for $P$-wave transitions between $s$-wave and
$s$-wave baryons \cite{Yan}, six couplings $h_{2}-h_7$ for the
$S$-wave transitions between $s$-wave and $p$-wave baryons, and
eight couplings $h_{8}-h_{15}$ for the $D$-wave transitions between
$s$-wave and $p$-wave baryons \cite{Pirjol}.

\subsection{Strong decays of $s$-wave charmed baryons}

In principle, the coupling $g_1$ can be determined from the decay
$\Sigma_c^*\to\Sigma_c\pi$. Unfortunately, this strong decay is
kinematically prohibited since the mass difference between
$\Sigma_c^*$ and $\Sigma_c$ is only of order 65 MeV. Consequently,
the coupling $g_1$ cannot be extracted directly from the strong
decays of heavy baryons. As for the coupling $g_2$, one can use the
measured rates of $\Sigma_c^{++}\to\Lambda_c^+\pi^+$,
$\Sigma_c^{*++}\to\Lambda_c^+\pi^+$ and
$\Sigma_c^{*0}\to\Lambda_c^+\pi^-$  as inputs to obtain
 \begin{eqnarray} \label{eq:g2}
 |g_2|=0.605^{+0.039}_{-0.043}\,,\quad 0.57\pm0.04\,, \quad
 0.60\pm0.04\,,
 \end{eqnarray}
respectively, where we have neglected the tiny contributions from
electromagnetic decays. Hence, the averaged $g_2$ is
 \begin{eqnarray}
 |g_2|=0.591\pm0.023\,.
 \end{eqnarray}
Using this value of $g_2$,  the predicted total width of
$\Xi_c^{*+}$ is found to be in the vicinity of the current limit
$\Gamma(\Xi_c^{*+})<3.1$ MeV \cite{CLEOb}.

It is clear from Table \ref{tab:strongdecayS} that  the strong
decay width of $\Sigma_c$ is smaller than that of $\Sigma_c^*$ by a
factor of $\sim 7$, although they will become the same in the limit
of heavy quark symmetry. This is ascribed to the fact that the c.m.
momentum of the pion is around 90 MeV in the decay
$\Sigma_c\to\Lambda_c\pi$ while it is two times bigger in
$\Sigma_c^*\to\Lambda_c\pi$. Since $\Sigma_c$ states are
significantly narrower than their spin-$\frac32$ counterparts, this
explains why the measurement of their widths came out much later.

\begin{table}[h]
\caption{Decay widths (in units of MeV) of $s$-wave charmed
baryons. } \label{tab:strongdecayS}
\begin{center}
\begin{tabular}{|c|c|c|} \hline \hline
~~~~~~~Decay~~~~~~~ & ~~~Expt.~~~ & ~HHChPT~ \\
\hline
 $\Sigma_c^{++}\to\Lambda_c^+\pi^+$ & $2.23\pm0.30$ & input  \\ \hline
 $\Sigma_c^{+}\to\Lambda_c^+\pi^0$ & $<4.6$ & $2.5\pm0.2$  \\ \hline
 $\Sigma_c^{0}\to\Lambda_c^+\pi^-$ & $2.2\pm0.4$ & input   \\ \hline
 $\Sigma_c(2520)^{++}\to\Lambda_c^+\pi^+$ & $14.9\pm1.9$ & input   \\  \hline
 $\Sigma_c(2520)^{+}\to\Lambda_c^+\pi^0$ & $<17$ & $16.6\pm1.3$  \\  \hline
 $\Sigma_c(2520)^{0}\to\Lambda_c^+\pi^-$ & $16.1\pm2.1$ & input   \\  \hline
 $\Xi_c(2645)^+\to\Xi_c^{0,+}\pi^{+,0}$ & $<3.1$ & $2.7\pm0.2$  \\  \hline
 $\Xi_c(2645)^0\to\Xi_c^{+,0}\pi^{-,0}$ & $<5.5$ & $2.8\pm0.2$  \\ \hline \hline
\end{tabular}
\end{center}
\end{table}

\subsection{Strong decays of $p$-wave charmed baryons}

\begin{table}[h]
\caption{Same as Table \ref{tab:strongdecayS} except for $p$-wave
charmed baryons. } \label{tab:strongdecayP}
\begin{center}
\begin{tabular}{|c|c|c|} \hline \hline
~~~~~~~Decay~~~~~~~ & ~~~Expt.~~~ & ~~HHChPT~~  \\
\hline
 $\Lambda_c(2595)^+\to (\Lambda_c^{+}\pi\pi)_R$ & $2.63^{+1.56}_{-1.09}$ & input \\ \hline
 $\Lambda_c(2595)^+\to \Sigma_c^{++}\pi^-$ & $0.65^{+0.41}_{-0.31}$ & $0.72^{+0.43}_{-0.30}$  \\ \hline
 $\Lambda_c(2595)^+\to \Sigma_c^{0}\pi^+$ & $0.67^{+0.41}_{-0.31}$ & $0.77^{+0.46}_{-0.32}$  \\ \hline
 $\Lambda_c(2595)^+\to \Sigma_c^{+}\pi^0$ & & $1.57^{+0.93}_{-0.65}$  \\ \hline
 $\Lambda_c(2625)^+\to \Sigma_c^{++}\pi^-$ & $<0.10$ & $\leq 0.029$  \\ \hline
 $\Lambda_c(2625)^+\to \Sigma_c^{0}\pi^+$ & $<0.09$ & $\leq 0.029$  \\ \hline
 $\Lambda_c(2625)^+\to \Sigma_c^{+}\pi^0$ & & $\leq 0.041$  \\ \hline
 $\Lambda_c(2625)^+\to \Lambda_c^+\pi\pi$ & $<1.9$ & $\leq 0.21$   \\ \hline
 $\Sigma_c(2800)^{++}\to\Lambda_c\pi,\Sigma_c^{(*)}\pi$ & $75^{+22}_{-17}$ & input \\ \hline
 $\Sigma_c(2800)^{+}\to\Lambda_c\pi,\Sigma_c^{(*)}\pi$ & $62^{+60}_{-40}$ & input  \\ \hline
 $\Sigma_c(2800)^0\to\Lambda_c\pi,\Sigma_c^{(*)}\pi$ & $61^{+28}_{-18}$ & input  \\ \hline
 $\Xi_c(2790)^+\to\Xi'^{0,+}_c\pi^{+,0}$ & $<15$ & $8.0^{+4.7}_{-3.3}$  \\ \hline
 $\Xi_c(2790)^0\to\Xi'^{+,0}_c\pi^{-,0}$  & $<12$ & $8.5^{+5.0}_{-3.5}$  \\ \hline
 $\Xi_c(2815)^+\to\Xi^{*+,0}_c\pi^{0,+}$ & $<3.5$ & $3.4^{+2.0}_{-1.4}$  \\ \hline
 $\Xi_c(2815)^0\to\Xi^{*+,0}_c\pi^{-,0}$ & $<6.5$ & $3.6^{+2.1}_{-1.5}$ \\ \hline \hline
\end{tabular}
\end{center}
\end{table}

Some of the $S$-wave and $D$-wave couplings of $p$-wave baryons to
$s$-wave baryons can be determined. In principle, the coupling
$h_2$ is readily extracted from $\Lambda_c(2595)^+\to
\Sigma_c^0\pi^+$ with $\Lambda_c(2595)$ being identified as
$\Lambda_{c1}(\frac12^-)$. However, since
$\Lambda_c(2595)^+\to\Sigma_c\pi$ is kinematically barely allowed,
the finite width effects of the intermediate resonant states could
become important \cite{Falk03}. Before proceeding to a more precise
determination of $h_2$, we make several remarks on the partial
widths of $\Lambda_c(2595)^+$ decays. (i) PDG \cite{PDG} has
assumed the isospin relation, namely,
$\Gamma(\Lambda_c^+\pi^+\pi^-)=2\Gamma(\Lambda_c^+\pi^0\pi^0)$ to
extract the branching ratios for $\Sigma_c\pi$ modes. However, the
decay $\Lambda_c(2595)\to\Lambda_c\pi\pi$ occurs very close to the
threshold as $m_{\Lambda_c(2595)}-m_{\Lambda_c}=308.9\pm0.6$ MeV.
Hence, the phase space is very sensitive to the small
isospin-violating mass differences between members of pions and
charmed Sigma baryon multiplets. Since the neutral pion is slightly
lighter than the charged one, it turns out that both
$\Lambda_c^+\pi^+\pi^-$ and $\Lambda_c^+\pi^0\pi^0$ have very
similar rates. (ii) Taking ${\cal B}(\Lambda_c(2595)^+\to
\Lambda_c^+\pi^+\pi^-)\approx 0.5$ and using the measured ratios of
$\Lambda_c(2595)^+\to \Sigma_c^{++}\pi^-)$ and $\Sigma_c^{0}\pi^+$
relative to $\Lambda_c(2595)^+\to \Lambda_c^+\pi^+\pi^-)$, we
obtain
 \begin{eqnarray}
 \Gamma(\Lambda_c(2595)^+\to\Sigma_c^{++}\pi^-) &=& 0.65^{+0.41}_{-0.31}\,{\rm MeV},
 \nonumber \\
 \Gamma(\Lambda_c(2595)^+\to\Sigma_c^{0}\pi^+) &=& 0.67^{+0.41}_{-0.31}\,{\rm
 MeV}\,.
 \end{eqnarray}
(iii) The non-resonant or direct three-body decay mode
$\Lambda_c^+\pi^+\pi^-$ has a branching ratio of $0.14\pm0.08$
\cite{PDG}. Assuming the same for $\Lambda_c^+\pi^0\pi^0$ and using
the measured total width of $\Lambda_c(2595)^+$, we are led to
 \begin{eqnarray}
 \Gamma(\Lambda_c(2595)^+\to\Lambda_c^+\pi\pi)_{\rm R} &=& (2.63^{+1.56}_{-1.09})\,{\rm
 MeV}, \nonumber \\
 \Gamma(\Lambda_c(2595)^+\to\Lambda_c^+\pi\pi)_{\rm NR} &=&
 (0.97^{+0.76}_{-0.64})\,{\rm
 MeV}.
 \end{eqnarray}

Consider the pole contributions to the decays
$\Lambda_c(2595)^+,\Lambda_c(2625)^+\to \Lambda_c^+\pi\pi$  with
the finite width effects included. The intermediate states of
interest are $\Sigma_c$ and $\Sigma_c^*$ poles.  The decay rates
depend on two coupling constants $h_2$ and $h_8$. Identifying the
calculated $\Gamma(\Lambda_c(2595)^+\to\Lambda_c^+\pi\pi)$ with the
resonant one, we find
 \begin{eqnarray}\label{eq:h2fw}
 |h_2|=0.437^{+0.114}_{-0.102}\,, \quad |h_8|< 3.65\times
 10^{-3}\,{\rm MeV}^{-1}.
 \end{eqnarray}
Assuming that the total width of $\Lambda_c(2593)^+$ is saturated
by the resonant $\Lambda_c^+\pi\pi$ 3-body decays, Pirjol and Yan
obtained $|h_2|=0.572^{+0.322}_{-0.197}$ and $|h_8|\leq
(3.50-3.68)\times 10^{-3}\,{\rm MeV}^{-1}$ \cite{Pirjol}. Our value
of $h_2$ is slightly smaller since in our case, the $\Sigma_c$ and
$\Sigma_c^*$ poles only describe the {\it resonant} contributions
to the total width of $\Lambda_c(2593)$.

The $\Xi_c(2790)$ and $\Xi_c(2815)$ baryons form a doublet
$\Xi_{c1}({1\over 2}^-,{3\over 2}^-)$. $\Xi_c(2790)$ decays to
$\Xi'_c\pi$, while $\Xi_c(2815)$ decays to $\Xi_c\pi\pi$,
resonating through $\Xi^*_c$, i.e. $\Xi_c(2645)$. Using the
coupling $h_2$ obtained from (\ref{eq:h2fw}) and the experimental
observation that the $\Xi_c\pi\pi$ mode in $\Xi_c(2815)$ decays is
consistent with being entirely via $\Xi^*_c\pi$
\cite{CLEO:Xic2815}, the predicted $\Xi_c(2790)$ and $\Xi_c(2815)$
widths are shown in Table \ref{tab:strongdecayP}. The predictions
are consistent with the current experimental limits.

Some information on the coupling $h_{10}$ can be inferred from the
strong decays of $\Sigma_c(2800)$. Assuming the widths of the
states $\Sigma_c(2800)^{++,+,0}$ are dominated by the two-body
$D$-wave modes $\Lambda_c\pi$, $\Sigma_c\pi$ and $\Sigma_c^*\pi$,
and applying the quark model relation $|h_8|=|h_{10}|$
\cite{Pirjol}, we then have
 \begin{eqnarray} \label{eq:h8}
|h_8|\leq (0.86^{+0.08}_{-0.10})\times 10^{-3}\,{\rm MeV}^{-1}\,,
 \end{eqnarray}
which improves the previous limit (\ref{eq:h2fw}) by a factor of 4.

\begin{acknowledgments}
I wish to thank Chun-Khiang Chua for collaboration on this
interesting subject and the organizers for organizing this very
stimulating workshop.
\end{acknowledgments}

\bigskip 

\end{document}